\documentclass[onecolumn,nobibnotes,nofootinbib,superscriptaddress]{revtex4}
\usepackage{amsmath,amssymb,bm}
\usepackage[a4paper,bindingoffset=0.2in,left=0.8in,right=0.8in,top=1in,bottom=1in,footskip=.25in]{geometry}
\usepackage{graphicx,subfigure,epsfig}
\usepackage{bigints}

\newcommand{\be}{\begin{equation}}
\newcommand{\ee}{\end{equation}}
\newcommand{\bea}{\begin{eqnarray}}
\newcommand{\eea}{\end{eqnarray}}
\newcommand{\benn}{\begin{displaymath}}
\newcommand{\eenn}{\end{displaymath}}
\newcommand{\beann}{\begin{eqnarray*}}
\newcommand{\eeann}{\end{eqnarray*}}

\newcommand{\nn}{\nonumber\\}

\newcommand{\abar}{\bar{\alpha}_s}

\begin{document}

\title{Solution to the Sudakov suppressed Balitsky-Kovchegov equation and its application to the HERA data}
\author{Wenchang Xiang}
\email{wxiangphy@gmail.com}
\affiliation{Guizhou Key Laboratory in Physics and Related Areas, and Guizhou Key Laboratory of Big Data Statistic Analysis, Guizhou University of Finance and Economics, Guiyang 550025, China}
\affiliation{Department of Physics, Guizhou University, Guiyang 550025, China}
\author{Mengliang Wang}
\email{mengliang.wang@mail.gufe.edu.cn}
\affiliation{Guizhou Key Laboratory in Physics and Related Areas, Guizhou University of Finance and Economics, Guiyang 550025, China}
\author{Yanbing Cai}
\email{myparticle@163.com}
\affiliation{Guizhou Key Laboratory in Physics and Related Areas, Guizhou University of Finance and Economics, Guiyang 550025, China}
\author{Daicui Zhou}
\email{dczhou@mail.ccnu.edu.cn}
\affiliation{Key Laboratory of Quark and Lepton Physics (MOE), and Institute of Particle Physics, Central China Normal University, Wuhan 430079, China}


\begin{abstract}
We analytically solve the Sudakov suppressed Balitsky-Kovchegov evolution equation with the fixed and running coupling constants in the saturation region. The analytic solution of the $S$-matrix shows the $\exp(\mathcal{O}(\eta^2))$ rapidity dependence of the solution with the fixed coupling constant is replaced by $\exp(\mathcal{O}(\eta^{3/2}))$ dependence in the smallest dipole running coupling case rather than obeying the law found in our previous publication, in which all the solutions of the next-to-leading order evolution equations comply with $\exp(\mathcal{O}(\eta))$ rapidity dependence once the QCD coupling is switched from the fixed coupling to the smallest dipole running coupling prescription. This finding indicates that the corrections of the sub-leading double logarithms in the Sudakov suppressed evolution equation are significant, which compensate part of the evolution decrease of the dipole amplitude made by running coupling effect. To test the analytic findings, we calculate the numerical solutions of the Sudakov suppressed evolution equation, the numerical results confirm the analytic outcomes. Moreover, we use the numerical solutions of the evolution equation to fit the HERA data. It shows that the Sudakove suppressed evolution equation can give good quality fit to the data.
\end{abstract}

\maketitle


\section{Introduction}
\label{sec:intro}
It is known that the energy evolution of the high energy dipole-hadron scattering amplitude is governed by the non-linear Balitsky-JIMWLK\footnote{The JIMWLK is the abbreviation of Jalilian-Marian, Iancu, McLerran, Weigert, Leonidov, Kovner.}\cite{B,JIMWLK1,JIMWLK2,JIMWLK3,JIMWLK4} hierarchy and its mean field approximation known as the Balitsky-Kovchegov (BK) equation\cite{B,K}. The BK equation is a closed equation which is convenient for the direct applications to phenomenological studies of saturation physics in available experimental data. However, the BK equation is a leading order (LO) evolution equation since it only resums leading logarithms~$\alpha_s\ln(1/x)$ arising from the successive emission of small-$x$ gluons, where the $\alpha_s$ is the QCD coupling and $x$ is the Bjorken variable. To give a realistic study of the observables, like reduced cross-section and structure function in deep inelastic scattering (DIS) at HERA, and forward particle production in high energy heavy ion collisions at RHIC and LHC, the next-to-leading order (NLO) corrections have to include into the high energy evolution equation of the dipole amplitude\cite{Xiang07,JS,JM,LM3,CX,Albacete17,YCai20}.

There are tremendous amount of efforts towards extending the LO evolution equation to the NLO accuracy in the literature\cite{Bnlo,KW,BC08,Beuf,NLOJIMWLK1,NLOJIMWLK2,NLOJIMWLK3,IMMST,LM15,LM16,Levin16,Xiang17,DIMST19,JZhou19,Xiang19}. A pioneer work on the NLO corrections to the BK equation was done by Balitsky in Ref.\cite{Bnlo} and Kovchegov-Weigert in Ref.\cite{KW}, in which the NLO corrections associated with the QCD coupling are resumed to all orders leading to the running coupling BK (rcBK) equation. The numerical studies of the rcBK equation found that the running coupling effect is large\cite{AK07}, and it is essential when the rcBK equation is used to quantitatively describe the structure functions measured at HERA\cite{JS,JM}. Although the rcBK equation gives a rather successful fit to the small-$x$ HERA data, the rcBK equation is partial of the full NLO evolution equation. From the Feynman diagram point of view, the rcBK equation only includes part of NLO corrections which refer to quark loop contributions. As we know that the gluon loops also contribute to the evolution of the dipole amplitude. The full NLO BK evolution equation which includes quark and gluon loops as well as tree gluon diagrams with quadratic and cubic nonlinearities, was derived by Balitsky and Chirilli in Ref.\cite{BC08}. The full NLO BK equation is so complicated that it was solved numerically seven years later than its derivation. Unfortunately, it was found that the solution of the full NLO BK equation is unstable\cite{LM15}. Mathematically, the reason for this difficulty was traced back to a large double transverse logarithmic correction in the evolution kernel of the full NLO BK equation. The physics behind the double logarithms (also called anti-collinear logarithms) is the time-ordering of the successive gluon emissions.

To cure the instability issues, one has to resum the radiative corrections enhanced by the double transverse logarithms to all orders. There are two approaches proposed to perform the resummations\cite{Beuf,IMMST}. One of the strategies for enforcing the time-ordering in the evolution with rapidity $Y$ of dipole projectile is to put kinematical constraints in the evolution kernel leading to a non-local equation in $Y$\cite{Beuf}. The other is to resum the double logarithmic corrections to all orders giving rise to a local collinearly improved Balitsky-Kovchegov (ciBK) equation in $Y$\cite{IMMST}. These two approaches are equivalent to each other in the leading double logarithmic level, although they bring the modifications to the structure of the evolution equation. Moreover, the condition of the time-ordering also takes the modifications to the corresponding initial conditions which are required for solving the evolution equations. However, the modifications of the initial conditions have been not properly implemented in the both approaches aforementioned, which lead to a matter of fact that the modifications in the initial condition impact not only on the evolution of the dipole amplitude in the region of low $Y$ but also on the asymptotic behavior in the region of large $Y$\cite{DIMST19}. This is an unexpected result, in fact that the high energy asymptotic behavior of the dipole amplitude should be inappreciably affected by the formulation of the initial condition.

To overcome the instability problems and solve it fundamentally, an effective method was proposed by the authors in Ref.\cite{DIMST19} inspired from previous experience on handling similar issues of the NLO BFKL\footnote{The BFKL is the abbreviation of Balitsky, Fadin, Kuraev, Lipatov.} equation\cite{Lipatov76,KLF77,BL78}. In Ref.\cite{DIMST19}, they used a new rapidity variable (rapidity of the target, $\eta$) instead of the rapidity of the projectile ($Y$) to be as the ``evolution time'', and reorganized the perturbative QCD theory for the evolution of the dipole amplitude. The advantage of this method is that the time-ordering condition is automatically satisfied, and then the anti-collinear contributions are absent in the target rapidity evolution. Furthermore, this choice of evolution variable is more reasonable, since the rapidity of the target is the one indeed used in the DIS rather than $Y$. A new version of non-local collinearly improved Balitsky-Kovchegov (non-local ciBK) equation was obtained\cite{DIMST19}, which was shown to give a rather good fits to the HERA data\cite{DIST20}. Soon after the non-local ciBK equation was established, it was found that there are important corrections to the evolution kernel from the sub-leading double logarithms located beyond the strong time-ordering region, it was shown that the sub-leading double logarithms arise from the incomplete cancellation between real and virtual corrections and are the typical Sudakov type ones\cite{JZhou19}. When these double logarithms are resummed to all orders, a Sudakov suppressed Balitsky-Kovchegov (SSBK) equation in the evolution of the target rapidity is obtained\cite{JZhou19}. The kernel of the SSBK equation is modified significantly by the sub-leading double logarithms.

In this paper, we shall solve analytically the SSBK and non-local ciBK equations in the saturation region with the smallest dipole running coupling prescription (SDRCP). Note that the reason why we use the SDRCP is that it was shown to be an effective QCD coupling in our previous publication\cite{Xiang20}. To see the variances in the solutions of two types of evolution equations (based on rapidity of projectile $Y$ and on rapidity of target $\eta$) due to the change of evolution variable from $Y$ to $\eta$, we firstly recall the analytic solutions of the LO BK, rcBK and ciBK equations in $Y$, and we shall use these solutions for the comparisons with the ones of the respective evolution in $\eta$. We find that the analytic solutions of the non-local ciBK and SSBK in $\eta$ with the fixed coupling Eqs.(\ref{solfixnon}) and (\ref{solSudBK}) are similar to the one gained at LO BK in $Y$ Eq.(\ref{SolLOBK}), except that the coefficients in the exponent are different. We also find that the solution of the non-local ciBK in $\eta$ with the SDRCP Eq.(\ref{solrcBK_eta}) is analogous to the one obtained at rcBK in $Y$ Eq.(\ref{solrcBK}). Surprisingly, the analytic solution of the SSBK equation with SDRCP shows that the rapidity in the exponent of the $S$-matrix has rapidity raised to the power of $3/2$ dependence, $\exp(\mathcal{O}(\eta^{3/2}))$, instead of linear rapidity dependence $\exp(\mathcal{O}(\eta))$, which do not obey the law found in our previous studies\cite{Xiang20}, in which we showed that the solutions of all kinds of NLO BK equations in $Y$ with SDRCP have linear rapidity dependence in the exponent of the $S$-matrix. Coincidentally, the rapidity dependence of the solution to the SSBK in $\eta$ with SDRCP is similar to the one obtained at the full NLO BK in $Y$, $\exp(\mathcal{O}(Y^{3/2}))$, with the parent dipole running coupling prescription (PDRCP)\cite{Xiang17,Xiang19,Xiang20}.

To test the analytic findings mentioned above, we numerically solve the SSBK and non-local ciBK equations with the fixed and running coupling constants, via focusing on the physics in the saturation region. The numerical results confirm our analytic findings, see Fig.\ref{figSSBK}. Furthermore, the SSBK equation is used to fit the HERA data. It shows that the theoretical calculations almost overlap with all the data points, see Fig.\ref{figred}, and a reasonable value of $\chi^2/\mathrm{d.o.f}=1.128$ is obtained from the fit, which indicate that the SSBK equation can give a rather good description to the data.

\section{The leading order, running coupling and collinearly improved evolution equations of color dipoles in $Y$-representation}
\label{sec:loandrc}
In this section, we give a brief review of the LO BK, rcBK, ciBK equations in $Y$-representation in order to collect the basic elements of the BK equations, which shall be useful for the latter discussion.
The review shall also give us a chance to introduce notations and explain the kinematics of color dipoles.

\subsection{The Balitsky-Kovchegov equation and its analytic solution at leading order}
\label{sec:lobk}

Consider a high energy dipole which is consisted of a quark-antiquark pair, scattering off a hadronic target, in the eikonal approximation the dipole scattering matrix can be written as a correlator of two Wilson lines\cite{BC08}
\be
S(\bm{x},\bm{y};Y)= \frac{1}{N_c}\big\langle \mathrm{Tr}\{U(\bm{x})U^{\dagger}(\bm{y})\}\big\rangle_Y,
\label{S_matrix}
\ee
where $\bm{x}$ and $\bm{y}$ are the transverse coordinates of the quark and antiquark, and $U$ is the time ordered Wilson line
\be
U(\bm{x})=\mathrm{P}\exp\Big[ig\int dx^{-}A^{+}(x^{-},\bm{x})\Big],
\ee
with $A^{+}(x^{-},\bm{x})$ to be as the gluon field of the target hadron. Note that the average in Eq.(\ref{S_matrix}) is given by the average over the target gluon field configurations at a fixed rapidity.

The rapidity evolution of the dipole scattering matrix can be described by the BK evolution equation\cite{B,K}
\be
\frac{\partial}{\partial Y} S(\bm{x},\bm{y};Y) =
           \int d^2\bm{z} K^\mathrm{LO}(\bm{x},\bm{y},\bm{z})
         \left [S(\bm{x},\bm{z}; Y)S(\bm{z},\bm{y}; Y) - S(\bm{x},\bm{y}; Y) \right],
\label{LOBK}
\ee
where the $K^\mathrm{LO}(\bm{x},\bm{y},\bm{z})$ is LO evolution kernel describing the dipole splitting probability density, and has form
\be
K^\mathrm{LO}(\bm{x},\bm{y},\bm{z})=\frac{\bar{\alpha}_s}{2 \pi} \frac{r^2}{r_1^2r_2^2},
\label{LOBK_K}
\ee
with coupling being redefined as $\bar{\alpha}_s = \alpha_sN_c/\pi$. The $\bm{r}=\bm{x}-\bm{y}$, $\bm{r}_1=\bm{x}-\bm{z}$ and $\bm{r}_2=\bm{z}-\bm{y}$ in Eq.(\ref{LOBK_K}) are the transverse sizes of the parent dipole and two emitted daughter dipoles, respectively. In the large $N_c$ limit, the Eq.(\ref{LOBK}) describes the evolution of the original dipole ($\bm{x}$, $\bm{y}$) splitting into two daughter dipoles, ($\bm{x}$, $\bm{z}$) and ($\bm{z}$, $\bm{y}$) sharing a common transverse coordinate of the emitted gluon $\bm{z}$. The non-linear term in $S$ on the right hand side (r.h.s) of Eq.(\ref{LOBK}) depicts the two daughter dipoles interaction with the target simultaneously, which is usually called as ``real'' term due to its really measurement of the scattering of the soft gluon. While the linear term in Eq.(\ref{LOBK}) is referred as ``virtual'' term since it gives the survival probability for the original dipole at the time of scattering. We would like to note that the BK equation is a LO evolution equation since it resums only the leading logarithmic $\alpha_s\ln(1/x)$ corrections in the fixed coupling case. Meanwhile, the BK equation is a mean field version of the Balitsky-JIMWLK hierarchy\cite{JIMWLK1,JIMWLK2,JIMWLK3,JIMWLK4} equations, in which higher order correlations are neglected.

Now, let us solve the BK equation analytically in the saturation region. In this region we know that the parton density in the target is so high that the interaction between the dipole and target is very strong, which leads to the scattering amplitude close to unit, namely $T\sim 1$. Via using the relation between the scattering amplitude and scattering matrix, $T=1-S$, one can get $S\sim0$. Therefore, one can neglect the contribution from the non-linear term, the Eq.(\ref{LOBK}) becomes
\be
\frac{\partial}{\partial Y} S(r, Y) \simeq
           -\int d^2r_1 K^{\mathrm{LO}}(r, r_1, r_2)
           S(r, Y).
\label{simLOBK}
\ee

To get the solution of the Eq.(\ref{simLOBK}), we need to know the upper and lower integral bounds. The lower integral bound can be set to $1/Q_s$ since the saturation condition requires that the size of the dipole should be larger than the typical transverse size $r_s\sim1/Q_s$. Here the $Q_s$ is the saturation scale which is an intrinsic momentum scale playing a role of the separation of dilute region from saturation region. The upper integral bound can be set to the original dipole size $r$ although few daughter dipoles may have size larger than $r$, however the evolution kernel is rapid decrease when $r_1(r_2)>r$. Hence, the contribution from those dipoles can be negligible. With the upper and lower bounds, the Eq.(\ref{LOBK}) can be rewritten as
\be
\frac{\partial}{\partial Y} S(r, Y) \simeq
           -\int^{r}_{1/Q_s} d^2r_1 K^{\mathrm{LO}}(r, r_1, r_2)
           S(r, Y).
\label{simLOBK1}
\ee
\begin{figure}[h!]
\setlength{\unitlength}{1.5cm}
\begin{center}
\epsfig{file=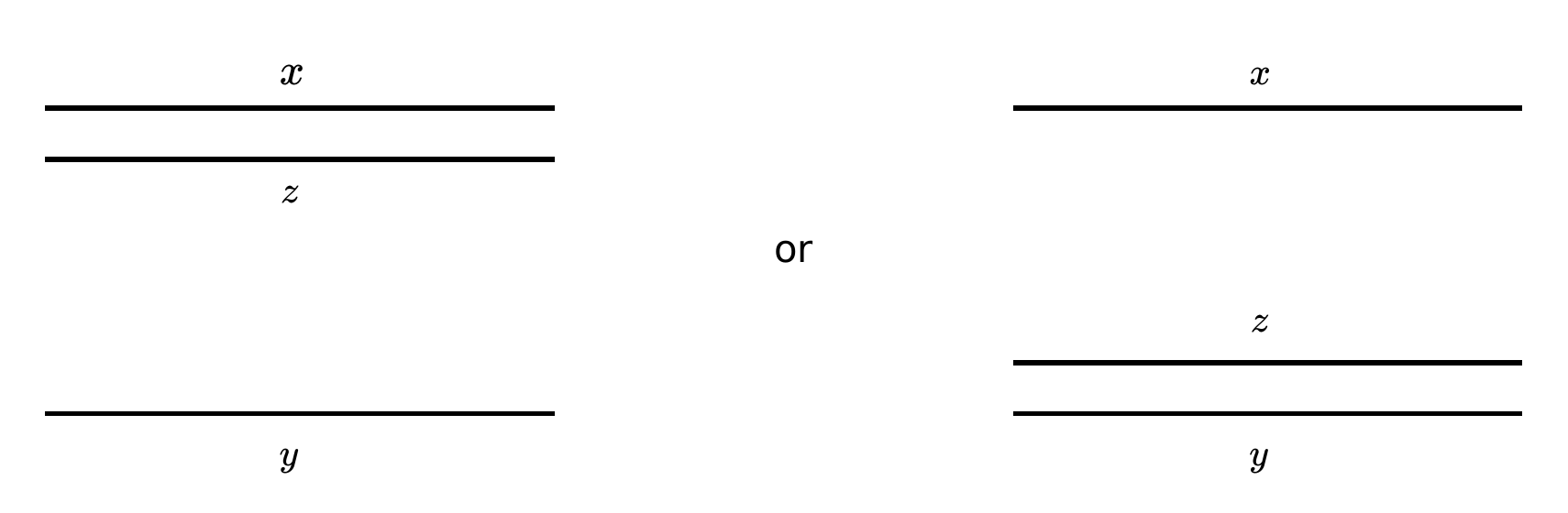, width=11cm,height=3.5cm}
\end{center}
\caption{The transverse coordinates of the parent and daughter dipoles in the saturation region.}
\label{figcoordinate}
\end{figure}

In the saturation region, one can find that the integral in Eq.(\ref{simLOBK1}) is governed by the region either from the transverse coordinate of the emitted gluon approaching to quark leg of the parent dipole, $1/Q_s\ll |\bm{r_1}|\ll |\bm{r}|$ and $|\bm{r_2}|\sim |\bm{r}|$, or the transverse coordinate of the emitted gluon approaching to antiquark leg of the parent dipole, $1/Q_s\ll |\bm{r_2}|\ll |\bm{r}|$ and $|\bm{r_1}|\sim |\bm{r}|$, see Fig.\ref{figcoordinate}. In this study, we work in the region $|\bm{r_2}|\sim |\bm{r}|$, the evolution kernel simplifies to
\be
K^{\mathrm{LO}}(r, r_1, r_2) \simeq \frac{\bar{\alpha}_s}{2 \pi} \frac{1}{r_1^2},
\label{simK}
\ee
and the Eq.(\ref{simLOBK1}) can be rewritten as
\be
\frac{\partial}{\partial Y} S(r, Y) \simeq
           -2\frac{\bar{\alpha}_s}{2\pi}\pi\int_{1/Q_s^2}^{r^2}dr_1^2 \frac{1}{r_1^2}
           S(r, Y),
\label{eq_LOBK_apf}
\ee
where the factor $2$ on the r.h.s comes from the symmetry of the aforementioned two integral regions. If one performs the integrals over variables $r_1$ and $Y$ in Eq.(\ref{eq_LOBK_apf}), one can get the analytic solution of the LO BK equation in the saturation region as
\cite{Levin-Tuchin,Mueller},
\be
S(r, Y)=\exp\left[-\frac{c\abar^2}{2}(Y-Y_0)^2\right]S(r, Y_0)=\exp\left[-\frac{\ln^2(r^2Q_s^2)}{2c}\right]S(r, Y_0),
\label{SolLOBK}
\ee
where the $c$ is a constant coming from the saturation momentum $Q_s^2(Y)=\exp\left[c\bar{\alpha}_s(Y-Y_0)\right]Q_s^2(Y_0)$ with $Q_s^2(Y_0)r^2=1$.
The solution in Eq.(\ref{SolLOBK}) has been firstly derived in Ref.\cite{Levin-Tuchin}, which is called as Levin-Tuchin formula. One can see that the exponent of the scattering matrix $S$ has a quadratic rapidity dependence, which leads to the scattering matrix too small when the rapidity is large. In terms of the relation between the scattering matrix $S$ and scattering amplitude $T$, $T=1 - S$, we know that the evolution speed of scattering amplitude is too fast, which renders the LO BK equation insufficiently to describe the experimental data from HERA\cite{JS,JM}. So, one has to take into account the NLO order corrections to the LO BK equation, such as running coupling effect which can bring modifications to the evolution kernel leading to reduce the evolution speed of the dipole amplitude.


\subsection{The Balitsky-Kovchegov equation and its analytic solution in the case of running coupling}
\label{sec:rcBK}
The strong coupling constant $\alpha_s$ in the LO BK equation (\ref{LOBK}) was assumed to be constant when the LO BK equation was derived, which makes the LO BK equation to be as a leading logarithm (LL) accuracy evolution equation. A naive way to promote it to include NLO correction is to replace $\alpha_s\rightarrow\alpha_s(r^2)$ in the LO BK equation\cite{Albacete04}
\be
K^\mathrm{PDRCP}(r,r_1,r_2)=\frac{\alpha_s(r^2)N_c}{2 \pi^2} \frac{r^2}{r_1^2r_2^2},
\label{rcPD}
\ee
where the argument of the coupling constant is the transverse size of the parent dipole. For the $\alpha_s$, we shall use the running coupling at one loop accuracy
\be
\alpha_s(r^2) = \frac{1}{b\ln\left(\frac{1}{r^2\Lambda^2}\right)},
\label{runningc}
\ee
with $b=(11N_c-2N_f)/12\pi$.

There is another running coupling prescription proposed recently in Refs.\cite{IMMST2,Cepila1,DIST20,Xiang20} where they found that using the size of the smallest dipole to be as the argument of the coupling constant is favored by the HERA data at a phenomenological level, which is referred to the SDRCP. In the case of the SDRCP, the kernel can be written as
\be
K^\mathrm{SDRCP}(r,r_1,r_2)=\frac{\alpha_s(r_{\mathrm{min}}^2)N_c}{2 \pi^2} \frac{r^2}{r_1^2r_2^2} ~~~\mathrm{with}~~~ r_\mathrm{min}=\mathrm{min}\{r,r_1,r_2\}.
\label{rcSD}
\ee
We would like to point out that all the following studies referred to running coupling in this paper shall use the SDRCP unless otherwise specified prescription. 

The Eq.(\ref{rcPD}) is a naive way to include the running coupling effect into the LO BK equation, which is insufficient. The real running coupling corrections to the LO BK equation were calculated by including the contribution from the quark bubbles in the gluon lines. The calculations have been performed with two prescriptions by Balitsky in Ref.\cite{Bnlo} and Kovchegov and Weigert in Ref.\cite{KW}. We don't go details of the derivations of the rcBK equation, and just quote the results here, since it is out of interest of present paper.

The rcBK equation reads
\be
\frac{\partial}{\partial Y} S(r, Y) = \int\,d^2 r_1
  \,K^{\mathrm{rc}}(r, r_1, r_2)
  \left[S(r_1, Y)\,S(r_2, Y)-S(r, Y)\right],
\label{rcBK}
\ee
where $K^{\mathrm{rc}}(r, r_1, r_2)$ is the running coupling evolution kernel. One can find that the rcBK equation Eq.(\ref{rcBK}) has the same structure as the LO BK equation Eq.(\ref{LOBK}) but the evolution kernel modified by the running coupling effect. There are two different (Balitsky and Kovchegov-Weigert) prescriptions for the kernel of rcBK equation as mentioned above. In the case of Balitsky prescription\cite{Bnlo}, the running coupling kernel can be written as
\be
K^{\mathrm{rcBal}}(r, r_1, r_2) = \frac{N_c\alpha_s(r^2)}{2\pi^2}\left[\frac{r^2}{r_1^2r_2^2} + \frac{1}{r_1^2}\left(\frac{\alpha_s(r_1^2)}{\alpha_s(r_2^2)}-1\right)
+ \frac{1}{r_2^2}\left(\frac{\alpha_s(r_2^2)}{\alpha_s(r_1^2)}-1\right)\right].
\label{Bal_rc}
\ee
Under the Kovchegov-Weigert prescription\cite{KW}, the running coupling kernel reads
\be
  K^{\mathrm{rcKW}}(r, r_1, r_2)=\frac{N_c}{2\pi^2}\left[
    \alpha_s(r_1^2)\frac{1}{r_1^2}-
    2\,\frac{\alpha_s(r_1^2)\,\alpha_s(r_2^2)}{\alpha_s(R^2)}\,\frac{
      {\bf r}_1\cdot {\bf r}_2}{r_1^2\,r_2^2}+
    \alpha_s(r_2^2)\frac{1}{r_2^2} \right],
\label{KW_rc}
\ee
with
\be
R^2(r, r_1, r_2)=r_1\,r_2\left(\frac{r_2}{r_1}\right)^
{\frac{r_1^2+r_2^2}{r_1^2-r_2^2}-2\,\frac{r_1^2\,r_2^2}{
      {\bf r}_1\cdot{\bf r}_2}\frac{1}{r_1^2-r_2^2}}.
\label{R}
\ee
Note that we have found an interesting result in our previous studies in Ref.\cite{Xiang}, in which the Balitsky and Kovchegov-Weigert kernels reduce to the same one under the saturation condition
\be
K^{\mathrm{rc}}(r, r_1, r_2) = \frac{N_c}{2\pi^2}\frac{\alpha_s(r_1^2)}{r_1^2},
\label{SimrcBKK}
\ee
which indicates that the running coupling kernel is independent of the choice of prescription in the saturation region.

To analytically solve the rcBK equation in the saturation region, we substitute the simplified kernel Eq.(\ref{SimrcBKK}) into Eq.(\ref{rcBK}), and get
\be
\frac{\partial}{\partial Y} S(r, Y) = \frac{N_c}{2\pi^2} \int\,d^2 r_1
   \frac{\alpha_s(r_1^2)}{r_1^2}
  \left[S(r_1, Y)\,S(r_2, Y)-S(r, Y)\right].
\label{rcBKf}
\ee
Under the saturation condition, one knows that the scattering matrix is small, therefore one can neglect the quadratic term of the $S$-matrix in Eq.(\ref{rcBKf}).
The Eq.(\ref{rcBKf}) reduces to
\be
\frac{\partial S(r,Y)}{\partial Y}\simeq- 2\int_{1/Q_s}^{r}\,d^2 r_1
  \,\frac{\bar{\alpha}_s(r_1^2)}{2\pi r_1^2}S(r, Y),
\label{SKW_sr}
\ee
where the upper and lower bounds are determined using the same way as the LO BK equation in section \ref{sec:lobk}, and the factor $2$ on the r.h.s of Eq.(\ref{SKW_sr}) results from the symmetry of the two integral regions, see Fig.\ref{figcoordinate}. Doing the integral over the variables $r_1$ and $Y$ in Eq.(\ref{SKW_sr}), one obtains the analytic solution of the rcBK equation\cite{Xiang}
\bea
S(r,Y)&=&\exp\left\{-\frac{N_c}{b\pi}(Y-Y_0)\left[\ln\left(\frac{\sqrt{c(Y-Y_0)}}{\ln\frac{1}{r^2\Lambda^2}}\right)-\frac{1}{2}\right]\right\}S(r, Y_0)\nn
&=&\exp\left\{-\frac{N_c}{bc_0\pi}\ln^2\frac{Q_s^2}{\Lambda^2}\left[\ln\left(\frac{\ln\frac{Q_s^2}{\Lambda^2}}{\ln\frac{1}{r^2\Lambda^2}}\right)-\frac{1}{2}\right]\right\}S(r, Y_0),
\label{solrcBK}
\eea
where the NLO saturation momentum is used
\be
\ln\frac{Q_s^2}{\Lambda^2} = \sqrt{c_0(Y-Y_0)} + \mathcal{O}(Y^{1/6}).
\label{SM}
\ee
In Eq.(\ref{solrcBK}), one can see that the exponent of the $S$-matrix has a linear rapidity dependence in the running coupling case, while the exponent of the $S$-matrix quadratically depends on the rapidity in the fixed coupling case, see Eq.(\ref{SolLOBK}). The change of rapidity of the $S$-matrix from quadratic to linear dependence implies that the evolution speed of the dipole amplitude is slowed down by the running coupling effect, which is in agreement with the theoretical expectations\cite{Bnlo,AK07}. In addition, the phenomenological studies of the HERA experimental data in Refs.\cite{JS,JM} showed that the rcBK equation takes a significant improvement of the description of the HERA data, which indicate that the NLO corrections are important.

\subsection{The collinearly improved Balitsky-Kovchegov equation and its analytic solution}
\label{subsec_NLL BK}
It is known that the rcBK equation includes only the contributions from the quark bubbles, while a full NLO evolution equation should consider contributions from the quark and gluon bubbles as well as from the tree gluon diagrams with quadratic and cubic nonlinearities. The authors in Ref.\cite{BC08} have performed a comprehensive derivation, and gotten a full NLO BK equation which includes all the NLO corrections just mentioned. However, it has been found that the full NLO BK equation is unstable, since the dipole amplitude resulting from the full NLO BK equation can decrease with increasing rapidity, it can even turn to a negative value\cite{LM15}. The reason for this instability is traced to a large contribution from a double-logarithm in the kernel of the full NLO BK equation\cite{LM15}. To solve the unstable problem, a novel method was devised in Ref.\cite{IMMST} to resum double transverse logarithms to all orders, they obtained a resummed BK equation which governs the evolution in the double logarithmic approximation (DLA). Soon after the DLA BK equation released, they found that the single transverse logarithms (STL) also have a large corrections to the BK equation. Combining the single and double logarithmic corrections together, they got a collinearly improved (ci) BK equation, which reads\cite{IMMST2}
\be
\frac{\partial S(r, Y)}{\partial Y} = \int d^2r_1K^{\mathrm{ci}}(r,r_1,r_2)[S(r_1, Y)S(r_2, Y) - S(r, Y)],
\label{ciBK}
\ee
where the collinearly improved evolution kernel is\cite{LM16}
\be
K^{\mathrm{ci}}(r,r_1,r_2) = \frac{\bar{\alpha}_s}{2\pi}\frac{r^2}{r_1^2r_2^2}K^{\mathrm{STL}}K^{\mathrm{DLA}}
\label{ci_kernel}
\ee
with
\be
K^{\mathrm{STL}} = \exp\bigg\{-\abar A_1\bigg|\ln\frac{r^2}{\mathrm{min}\{r_1^2, r_2^2\}}\bigg|\bigg\}
\label{STLK}
\ee
and
\be
K^{\mathrm{DLA}} = \frac{J_1\bigg(2\sqrt{\abar \rho^2}\bigg)}{\sqrt{\abar\rho^2}}.
\label{DLAK}
\ee
The constant $A_1=11/12$ in Eq.(\ref{STLK}) is the DGLAP anomalous dimension, and the $J_1$ in Eq.(\ref{DLAK}) is the Bessel function of the first kind with
\be
\rho=\sqrt{\ln \Big(\frac{r_1^2}{r^2}\Big)\ln \Big(\frac{r_2^2}{r^2}\Big)}.
\label{rho}
\ee
We would like to point out when $\ln r_1^2/r^2\ln r_2^2/r^2 < 0$, then an absolute value is used and the Bessel function of the first kind $J_1$ changes to the modified Bessel function of the first kind $I_1$\cite{IMMST2}.

To analytically solve Eq.(\ref{ciBK}) in the saturation region, one needs to employ the saturation condition which means the scattering matrix is very small. Thus, one can neglect the quadratic term in Eq.(\ref{ciBK}) and keeps only the linear term, then the Eq.(\ref{ciBK}) becomes
\be
\frac{\partial S(r, Y)}{\partial Y} \simeq - 2\int_{1/Q_s}^{r} d^2r_1\frac{\bar{\alpha}_s}{2\pi}K^{\mathrm{CI}}S(r, Y),
\label{fnlobk_rsum_f}
\ee
where the upper and lower bounds of the integral are determined by using the same way as the ones in section \ref{sec:lobk}. As it was done in previous subsections, we work in the regime, $1/Q_s\ll|\bm{r}_1|\ll |\bm{r}|,~|\bm{r}_2|\sim|\bm{r}|$. So, we have $\ln(r^2_2/r^2)\simeq0$ leading to $\rho=0$, which implies the DLA kernel $K^{\mathrm{DLA}}\simeq 1$. This outcome confirms the statement that the double logarithm only plays a significant role in the week scattering phase-space\cite{IMMST}. Under the approximation just mentioned above, one has
\be
\frac{\partial S(r, Y)}{\partial Y} \simeq -2\int^{r}_{1/Q_s} d^2r_1\frac{\bar{\alpha}_s}{2\pi}\frac{r^2}{r_1^2r_2^2}\Bigg[\frac{r^2}{\mathrm{min}(r_1^2,r_2^2)}\Bigg]^{\pm\bar{\alpha}_sA_1}S(r, Y),
\label{ciBKfl}
\ee
whose solution is
\bea
S(r, Y) &=& \exp\Bigg\{-\frac{N_c}{2b^2\pi}(Y-Y_0)\bigg[\frac{3N_cA_1-b\pi}{\pi}+\frac{b\pi-N_cA_1}{\pi}\ln\bigg(\frac{c_0(Y-Y_0)}{\ln^2\frac{1}{r^2\Lambda^2}}\bigg)\bigg]\nn
&&\hspace{1.0cm}-\frac{2N_c^2A_1}{b^2\pi^2c_0}\sqrt{c_0(Y-Y_0)}\ln r^2\Lambda^2\Bigg\}S(r, Y_0)\nn
&=& \exp\Bigg\{-\frac{N_c}{2b^2\pi c_0}\ln^2\frac{Q_s^2}{\Lambda^2}\bigg[\frac{3N_cA_1-b\pi}{\pi}+\frac{b\pi-N_cA_1}{\pi}\ln\bigg(\frac{\ln^2\frac{Q_s^2}{\Lambda^2}}{\ln^2\frac{1}{r^2\Lambda^2}}\bigg)\bigg]\nn
&&\hspace{1.0cm}-\frac{2N_c^2A_1}{b^2\pi^2c_0}\ln\frac{Q_s^2}{\Lambda^2}\ln r^2\Lambda^2\Bigg\}S(r, Y_0).
\label{solciBK}
\eea
Note that the dominant terms in the exponents of Eqs.(\ref{solciBK}) and (\ref{solrcBK}) have linear rapidity dependence once the SDRCP is applied, which is a law found in Ref.\cite{Xiang20}. This outcome implies that the running coupling correction is a dominant effect over all the aforementioned NLO corrections, like resummations of double and single transverse logarithms, in the suppression of dipole evolution.

\section{The non-local next-to-leading order and Sudakov suppressed evolution equations of color dipoles in $\eta$-representation}
\label{sec:BKinEta}
In the previous section, all the BK equations are derived by following the evolution in terms of the rapidity of the projectile ($Y$). However, the recent studies in Ref.\cite{DIMST19} found that the NLO BK equation in $Y$ must to be re-established according to the rapidity of the dense target ($\eta$), since the evolution of the rapidity $Y$ could cause the instability to the NLO BK equation. In this section, we shall discuss the non-local ciBK equation in $\eta$ and its extended version which includes sub-leading double logarithms from the region beyond the strong time-ordering\cite{JZhou19}. The evolution equations in $\eta$-representation shall be solved analytically in the fixed and running coupling cases in the saturation region, respectively. The results are compared to the solutions obtained in the $Y$-representation.

\subsection{The non-local collinearly improved BK equation in $\eta$ and its analytic solution}
\label{sec:etarptt}
The non-local ciBK equation was reformulated via change of variables $\eta\equiv Y-\rho$ to transform the BK equation from the $Y$-representation to the $\eta$-representation,
\be
\bar{S}(r,\eta) \equiv S(r, Y=\eta+\rho),
\ee
where $\rho=\ln1/(r^2Q_0^2)$. The non-local ciBK equation can be written as\cite{DIMST19}

\bea
\frac{\partial \bar{S}(r,\eta)}{\partial\eta} &=& \int d^2r_1\frac{\bar{\alpha}_s}{2\pi}\frac{r^2}{r_1^2r_2^2}\Theta(\eta-\delta_{r,r_1,r_2})\big[\bar{S}(r_1,\eta-\delta_{r_1,r})\bar{S}(r_2,\eta-\delta_{r_2,r})-\bar{S}(r,\eta)\big]\nn
&&-\int d^2r_1\frac{\bar{\alpha}_s^2}{4\pi}\frac{r^2}{r_1^2r_2^2}\ln\frac{r_2^2}{r^2}\ln\frac{r_1^2}{r^2}\big[\bar{S}(r_1,\eta)\bar{S}(r_2,\eta)-\bar{S}(r,\eta)\big]\nn
&&+\int d^2r_1d^2r'_1\frac{\bar{\alpha}_s^2}{2\pi^2}\frac{r^2}{r'^2{r'_1}^2r_2^2}\Big[\ln\frac{{r'_2}^2}{r^2}+\delta_{r'_2,r}\Big]\bar{S}(r'_1,\eta)\big[\bar{S}(r',\eta)\bar{S}(r_2,\eta)-\bar{S}(r'_2,\eta)\big]\nn
&&+\bar{\alpha}_s^2\times ``\mathrm{regular}",
\label{nonlocalBK}
\eea
where the rapidity shifts $\delta_{r,r_1,r_2}$ and $\delta_{r_1,r}$ are defined as
\be
\delta_{r,r_1,r_2} = \mathrm{max}\Bigg\{0,\ln\frac{r^2}{\mathrm{min}\{r_1^2,r_2^2\}}\Bigg\},
\ee
and
\be
\delta_{r_1,r} = \mathrm{max}\Bigg\{0,\ln\frac{r^2}{r_1^2}\Bigg\},
\ee
and similarly for $\delta_{r_2,r}$. Note that on the r.h.s of Eq.(\ref{nonlocalBK}) there are only two NLO terms explicit display. All the other NLO terms are collectively denoted as ``regular''.
The roles of the rapidity shift $\delta_{r,r_1,r_2}$ as well as the step function in Eq.(\ref{nonlocalBK}) are to introduce constraints on the soft-to-hard evolution\cite{DIMST19}.

Now let's turn to solve the non-local ciBK equation analytically in the saturation region. As we know from the previous integrals in Eqs.(\ref{simLOBK1}) and (\ref{SKW_sr}), that the integral in Eq.(\ref{nonlocalBK}) is governed by the region either $1/\tilde{Q}_s\ll |\bm{r_1}|\ll |\bm{r}|$ and $|\bm{r_2}|\sim |\bm{r}|$, or $1/\tilde{Q}_s\ll |\bm{r_2}|\ll |\bm{r}|$ and $|\bm{r_1}|\sim |\bm{r}|$, where $\bar{Q}_s$ takes the role of saturation scale
\be
\bar{Q}_s^2(\eta)=\exp\big[\bar{c}\bar{\alpha}_s(\eta-\eta_0)\big]\bar{Q}_s^2(\eta_0).
\ee
We choose to work in the former region, where the $S$-matrix is negligibly small, thus we can neglect the quadratic term in Eq.(\ref{nonlocalBK}). In addition, $\ln(r_2^2/r^2)\sim0$ leads to the second term on the r.h.s of Eq.(\ref{nonlocalBK}) negligible. The non-local ciBK equation reduces to
\be
\frac{\partial \bar{S}(r,\eta)}{\partial\eta} \simeq -2\int_{1/\bar{Q}_s}^{r} d^2r_1\frac{\bar{\alpha}_s}{2\pi}\frac{1}{r_1^2}\Theta(\eta-\delta_{r,r_1,r_2})\bar{S}(r,\eta).
\label{simp-nonlocalBK}
\ee
%
\subsubsection{Solution with the fixed coupling constant}
In the fixed coupling case, we know that the QCD coupling can be viewed as a constant. Thus, the $\bar{\alpha}_s$ in Eq.(\ref{simp-nonlocalBK}) can be factorized out of the integral,
\be
\frac{\partial \bar{S}(r,\eta)}{\partial\eta} \simeq -\frac{\bar{\alpha}_s}{\pi}\int_{1/\bar{Q}_s}^{r} d^2r_1\frac{1}{r_1^2}\bar{S}(r,\eta),
\label{fix-nonlocalBK}
\ee
whose solution is\cite{DIMST19}
\be
\bar{S}(r,\eta) = \exp\bigg[-\frac{\bar{c}\abar^2}{2}(\eta-\eta_0)^2\bigg]\bar{S}(r,\eta_0) = \exp\bigg[-\frac{\ln^2(r^2\bar{Q}_s^2)}{2\bar{c}}\bigg]\bar{S}(r,\eta_0).
\label{solfixnon}
\ee
If one compares Eq.(\ref{solfixnon}) with Eq.(\ref{SolLOBK}), it is easy to find that both of them have quadratic rapidity dependence except for different coefficients in the exponent, although they are in different rapidity representation. It is known that the value of $\bar{c}$ is smaller than $c$\cite{DIMST19}, so the $S$-matrix in the $\eta$-representation is larger than the one in the $Y$-representation for the same value of rapidity.

\subsubsection{solution with the running coupling constant}
In the running coupling case, the QCD coupling is a function of the smallest dipole size. The coupling constant in Eq.(\ref{simp-nonlocalBK}) cannot be factorized out of the integral, the evolution equation (\ref{simp-nonlocalBK}) becomes
\be
\frac{\partial \bar{S}(r,\eta)}{\partial\eta} \simeq -\frac{N_c}{b\pi^2}\int_{1/\bar{Q}_s}^{r} d^2r_1\frac{1}{r_1^2\ln\frac{1}{r_1^2\Lambda^2}}\bar{S}(r,\eta),
\label{simp-nonlocalBK1}
\ee
whose solution is
\bea
\bar{S}(r,\eta)&=& \exp\left\{-\frac{N_c}{b\pi}(\eta-\eta_0)\left[\ln\left(\frac{\sqrt{\bar{c}_0(\eta-\eta_0)}}{\ln\frac{1}{r^2\Lambda^2}}\right)-\frac{1}{2}\right]\right\}S(r, \eta_0)\nn
&=&\exp\left\{-\frac{N_c}{b\bar{c}_0\pi}\ln^2\frac{\bar{Q}_s^2}{\Lambda^2}\left[\ln\left(\frac{\ln\frac{\bar{Q}_s^2}{\Lambda^2}}{\ln\frac{1}{r^2\Lambda^2}}\right)-\frac{1}{2}\right]\right\}S(r, \eta_0),
\label{solrcBK_eta}
\eea
where the NLO saturation momentum is used
\be
\ln\frac{\bar{Q}_s^2}{\Lambda^2} = \sqrt{\bar{c}_0(\eta-\eta_0)} + \mathcal{O}(\eta^{1/6}).
\label{SM_eta}
\ee
Note that the Eq.(\ref{solrcBK_eta}) has similar rapidity dependence as the Eqs.(\ref{solrcBK}) and (\ref{solciBK}) except for different coefficients in the exponent, although they are in different rapidity representation. This result means that the solution of the non-local ciBK equation also complies in the law what we found in Ref.\cite{Xiang20}. However, one shall find in the next subsection that the solution of the evolution equation including the corrections of the sub-leading double logarithms violates the law mentioned above.

\subsection{The Sudakov suppressed BK equation and its analytic solution}
\label{sec:sudBK}
It has been shown in Ref.\cite{JZhou19} that there are significant corrections coming from the regime beyond the strong ordering region, where the sub-leading double logarithms are induced due to the incomplete cancellation between the real corrections and virtual corrections. These double logarithms have typical Sudakov feature and can be resummed into an exponential type resulting in a Sudakov suppressed Balitsky-Kovchegov equation\cite{JZhou19}:
\be
\frac{\partial S(r, \eta)}{\partial \eta} = \int d^2r_1 K^{\mathrm{SS}}(r,r_1,r_2)[S(r_1, \eta)S(r_2, \eta) - S(r, \eta)],
\label{sudBK}
\ee
where the Sudakov suppressed evolution kernel is
\be
K^{\mathrm{SS}}(r,r_1,r_2) = \frac{\bar{\alpha}_s}{2\pi}\frac{r^2}{r_1^2r_2^2}\frac{1}{2}\Bigg\{\exp\bigg[-\frac{\bar{\alpha}_s}{2}\ln^2\frac{r^2}{r_1^2}\bigg]+\exp\bigg[-\frac{\bar{\alpha}_s}{2}\ln^2\frac{r^2}{r_2^2}\bigg]\Bigg\}.
\label{sudK}
\ee
\subsubsection{Solution with the fixed coupling constant}
\label{subsec_solfcc}
In the fixed coupling case, one can set the QCD coupling in Eq.(\ref{sudK}) to be as a constant. Thus, one can factorize it out of the integral during the analytic solving Eq.(\ref{sudBK}) in the saturation region. As it was done in the previous section, we need to use the saturation condition which indicates that the $S$-matrix is very small, the non-linear term on the r.h.s of Eq.(\ref{sudBK}) can be neglected. The SSBK equation becomes a linear evolution equation in $S$,
\be
\frac{\partial S(r, \eta)}{\partial \eta} \simeq -2\frac{\bar{\alpha}_s}{2\pi}\int^{r}_{1/\bar{Q}_s} d^2r_1 \frac{1}{r_1^2}\frac{1}{2}\Bigg\{\exp\bigg[-\frac{\bar{\alpha}_s}{2}\ln^2\frac{r^2}{r_1^2}\bigg] + 1\Bigg\}S(r, \eta),
\label{sudBKsimp}
\ee
where the upper and lower bounds are determined by the same way as the ones in the previous section. The factor 2 on the r.h.s of Eq.(\ref{sudBKsimp}) comes from the symmetry of the two integral regions, see Fig.\ref{figcoordinate}. We perform the integral over $r_1$ and $Y$ in Eq.(\ref{sudBKsimp}), and get its solution as
\bea
S(r, \eta) &=& \Bigg\{\exp\bigg[-\frac{\sqrt{\pi}}{8\bar{c}}\ln^2r^2\bar{Q}_s^2\bigg] + \exp\bigg[-\frac{1}{4\bar{c}}\ln^2r^2\bar{Q}_s^2\bigg]\Bigg\}S(r, \eta_0) \nn
&=& \Bigg\{\exp\bigg[-\frac{\sqrt{\pi}\bar{c}}{8}\bar{\alpha}_s^2(\eta-\eta_0)^2\bigg] + \exp\bigg[-\frac{\bar{c}}{4}\bar{\alpha}_s^2(\eta-\eta_0)^2\bigg]\Bigg\}S(r, \eta_0),
\label{solSudBK}
\eea
where we have used $\ln r^2\bar{Q}_s^2=\bar{c}\bar{\alpha}_s(\eta-\eta_0)$ with $\bar{c}$ to be as a constant. If one compares the solution of the SSBK equation in $\eta$ (\ref{solSudBK}) with the solution of the LO BK equation in $Y$ (\ref{SolLOBK}), it is easy to find that the $S$-matrix in the Sudakov case is larger than the LO one, since the exponential factor in the second term of Eq.(\ref{SolLOBK}) on the r.h.s is almost twice as large as the one in Eq.(\ref{solSudBK}). In other words, the scattering amplitude $T$ in the Sudakov case becomes smaller than the LO one, which indicates that the evolution speed of the dipole amplitude is slowed down by the Sudakov effect as compared to the LO one.

\subsubsection{Solution with the running coupling constant}
\label{subsec_solrcc}
In the running coupling case, one cannot factorize the QCD coupling out of the integral in Eq.(\ref{sudBK}), since the QCD coupling could be a function of integral variable. As it was done in the fixed coupling case, the Eq.(\ref{sudBK}) is solved in the saturation region, therefore we can neglect the non-linear term and just keep the linear term in $S$, the Eq.(\ref{sudBK}) simplifies to
\be
\frac{\partial S(r, \eta)}{\partial \eta} \simeq -2\int^{r}_{1/\bar{Q}_s} d^2r_1 \frac{\bar{\alpha}_s(r^2_{\mathrm{min}})}{2\pi}\frac{1}{r_1^2}\frac{1}{2}\Bigg\{\exp\bigg[-\frac{\bar{\alpha}_s}{2}\ln^2\frac{r^2}{r_1^2}\bigg] + 1\Bigg\}S(r, \eta).
\label{sudBKsimprc}
\ee
Note that the kernel in Eq.(\ref{sudBKsimprc}) is simplified by the fact that the integral is governed by the region either from the transverse coordinate of the emitted gluon approaching to quark leg of the parent dipole, $1/\bar{Q}_s\ll |\bm{r_1}|\ll |\bm{r}|$ and $|\bm{r_2}|\sim |\bm{r}|$, or the transverse coordinate of the emitted gluon approaching to antiquark leg of the parent dipole, $1/\bar{Q}_s\ll |\bm{r_2}|\ll |\bm{r}|$ and $|\bm{r_1}|\sim |\bm{r}|$, see Fig.\ref{figcoordinate}. The factor 2 on the r.h.s of Eq.(\ref{sudBKsimprc}) accounts for the symmetry of the integral regions just mentioned. We choose to work in the region, $1/\bar{Q}_s\ll |\bm{r_1}|\ll |\bm{r}|$ and $|\bm{r_2}|\sim |\bm{r}|$, the evolution kernel becomes
\bea
K^{\mathrm{SS}}(r,r_1,r_2) &=& \frac{\bar{\alpha}_s(r^2_\mathrm{min})}{2\pi}\frac{r^2}{r_1^2r_2^2}\frac{1}{2}\Bigg\{\exp\bigg[-\frac{\bar{\alpha}_s}{2}\ln^2\frac{r^2}{r_1^2}\bigg]+\exp\bigg[-\frac{\bar{\alpha}_s}{2}\ln^2\frac{r^2}{r_2^2}\bigg]\Bigg\}\nn
&\simeq& \frac{\bar{\alpha}_s(r^2_\mathrm{min})}{2\pi}\frac{1}{r_1^2}\frac{1}{2}\Bigg\{\exp\bigg[-\frac{\bar{\alpha}_s}{2}\ln^2\frac{r^2}{r_1^2}\bigg] + 1\Bigg\}.
\label{sudK1}
\eea
Performing the integral over variables $r_1$ and $\eta$ in Eq.(\ref{sudBKsimprc}), one can get its solution as
\bea
S(r, \eta) &=& \Bigg\{\exp\Bigg[-\frac{N_c}{4b\bar{c}_0\sqrt{\pi}\ln\frac{1}{r^2\Lambda^2}}\bigg(\frac{2}{3}\ln^3\frac{\bar{Q}_s^2}{\Lambda^2}+\ln\frac{1}{r^2\Lambda^2}\ln^2\frac{\bar{Q}_s^2}{\Lambda^2}\bigg)\Bigg]\nn
&&\hspace{0.5cm} + \exp\left[-\frac{N_c}{b\bar{c}_0\pi}\ln^2\frac{\bar{Q}_s^2}{\Lambda^2}\left(\ln\left(\frac{\ln\frac{\bar{Q}_s^2}{\Lambda^2}}{\ln\frac{1}{r^2\Lambda^2}}\right)-\frac{1}{2}\right)\right]\Bigg\}S(r,\eta_0)\nn
&=& \Bigg\{\exp\Bigg[-\frac{N_c}{4b\bar{c}_0\sqrt{\pi}\ln\frac{1}{r^2\Lambda^2}}\bigg(\frac{2}{3}\big(\bar{c}_0(\eta-\eta_0)\big)^{\frac{3}{2}}+\ln\frac{1}{r^2\Lambda^2}\bar{c}_0(\eta-\eta_0)\bigg)\Bigg]\nn
&&\hspace{0.5cm} + \exp\left[-\frac{N_c}{2b\pi}\bar{c}_0(\eta-\eta_0)\left(\ln\left(\frac{\sqrt{\bar{c}_0(\eta-\eta_0)}}{\ln\frac{1}{r^2\Lambda^2}}\right)-\frac{1}{2}\right)\right]\Bigg\}S(r,\eta_0),
\label{solSudBKrc}
\eea
where the saturation momentum in the NLO case is used, see Eq.(\ref{SM_eta}). The solution in Eq.(\ref{solSudBKrc}) deserves several important comments which are as follows:
\begin{itemize}
  \item If one compares the running coupling solution of the SSBK Eq.(\ref{solSudBKrc}) with the fixed coupling solution of the SSBK Eq.(\ref{solSudBK}), one can see that the rapidity dependence of the dominant term in the exponent changes from the quadratic rapidity dependence Eq.(\ref{solSudBK}) to the rapidity raised to power of 3/2 dependence Eq.(\ref{solSudBKrc}), rather than linear dependence. This result does not coincide with the law what we found in Ref.\cite{Xiang20} in which the rapidity dependence of the solutions of all the NLO evolution equations have linear dependence once the SDRCP is applied. This outcome indicates that the sub-leading double logarithms compensate part of decrease made by the running coupling effect.
  \item One can see that the solutions of the fixed coupling Eq.(\ref{solSudBK}) and running coupling Eq.(\ref{solSudBKrc}) SSBK equations have two terms due to the kernel having two terms, see Eq.(\ref{sudK}). The second terms in Eqs.(\ref{solSudBK}) and (\ref{solSudBKrc}) are similar as the respective ones in the non-local ciBK cases except an additional factor $1/2$ difference in the exponents, which implies that the first terms in Eqs.(\ref{solSudBK}) and (\ref{solSudBKrc}) could come from the incomplete cancellation of the real and virtual corrections.
  \item Interestingly, we find that the dominant term (the first term on the r.h.s of Eq.(\ref{solSudBKrc})) of the solution of the SSBK equation with SDRCP $\exp(\mathcal{O}(\eta^{3/2}))$ has similar rapidity dependence as the one obtained by solving the full NLO BK equation in $Y$ with the PDRCP $\exp(\mathcal{O}(Y^{3/2}))$.
\end{itemize}
To conclude, one can see that the sub-leading double logarithms resulting from the incomplete cancellation between the real and virtual corrections beyond the strong time-ordering region make significant change to the rapidity dependence of the dipole amplitude not only in fixed coupling case but also in the running coupling case.

\section{Numerical solutions of the Sudakov suppressed Balitsky-Kovchegov equation}
\label{sec:numsol}
In this section, we shall use the numerical method to solve the SSBK equation in order to test their analytic solutions obtained in the above section, especially focusing on the comparison of the numerical results located in the saturation region with the analytic solutions. As one knows that the dipole evolution equations are a set of complicated integro-differential equations, the Runge-Kutta method is needed to solve them on the lattice. The integrals in these equations are performed by the adaptive integration routines. In addition, the interpolation is needed during the numerical calculations, since some data points not locating on the lattice should be estimated. So, the cubic spline interpolation method is employed in this study. To simplify the computation, we employ the translational invariant approximation and assume the $S$-matrix independent of the impact parameter of the collisions, $S=S(|r|, Y)$. In terms of the above discussion, we choose to use the GNU Scientific Library (GSL) to perform the numerical computation, since the GSL includes almost all the functions required by the numerical solution to the evolution equations.

In order to solve the SSBK evolution equations, the McLerran-Venugopalan (MV) model is used to be as the initial condition\cite{MV},
\be
S^{\mathrm{MV}}(r, \eta=0) = \exp\bigg[-\bigg(\frac{r^2\bar{Q}_{s0}^2}{4}\bigg)^{\gamma}\ln\Big(\frac{1}{r^2\Lambda^2}+e\Big)\bigg],
\label{IC}
\ee
where we set $\bar{Q}^2_{s0}=0.15~\mathrm{GeV}^2$ at $\eta=0$, $\gamma=1$, and $\Lambda=0.2~\mathrm{GeV}$ for simplicity, but the $\bar{Q}^2_{s0}$ and $\gamma$ shall be free parameters when they use to fit the HERA data in the next section. The one-loop running coupling with $N_f=3$ and $N_c=3$, Eq.(\ref{runningc}), is used to be as the QCD coupling in this numerical simulation. In order to regularize the infrared behavior, the coupling value is freezed to $\alpha_s(r_{\mathrm{fr}})=0.75$ when $r>r_{\mathrm{fr}}$.

\begin{figure}[h!]
\setlength{\unitlength}{1.5cm}
\begin{center}
\epsfig{file=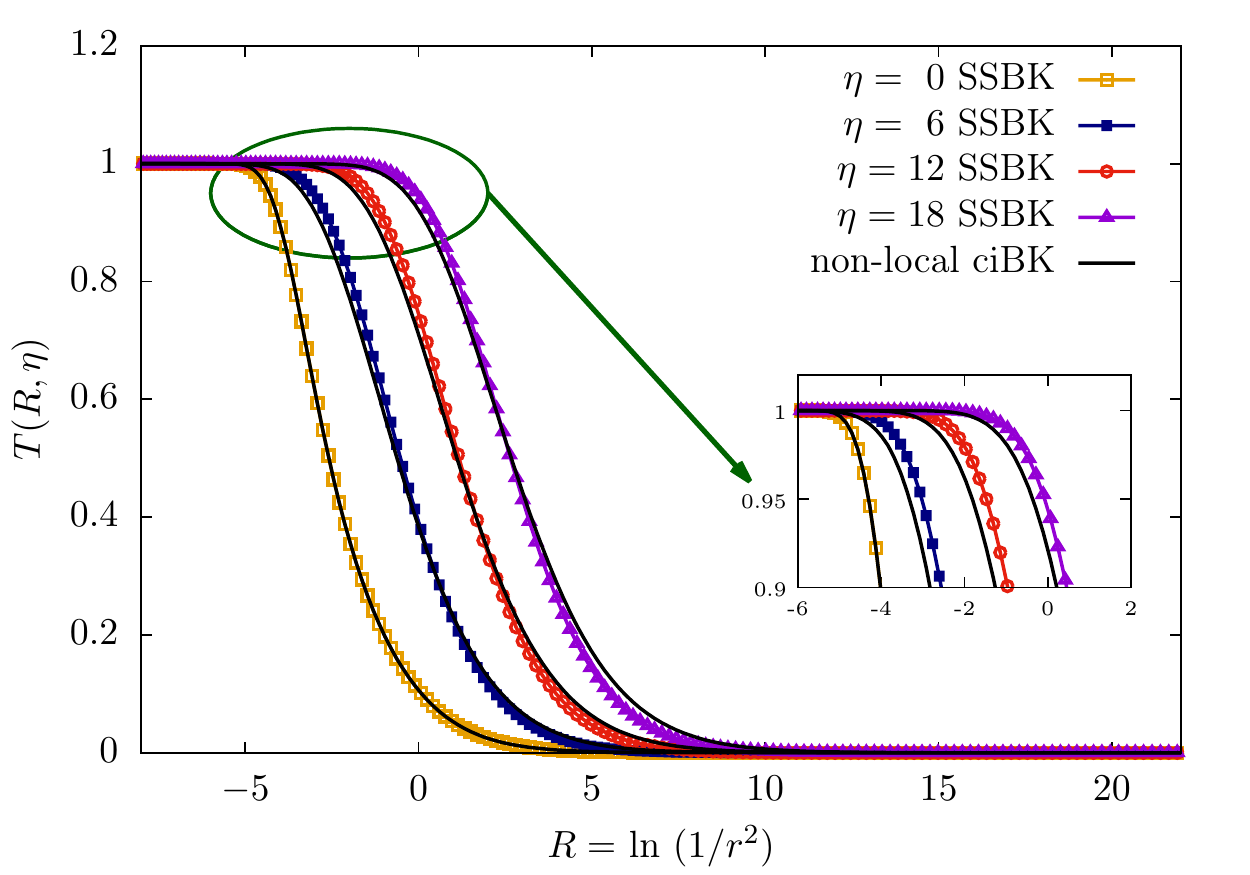, width=8cm,height=6cm}
\epsfig{file=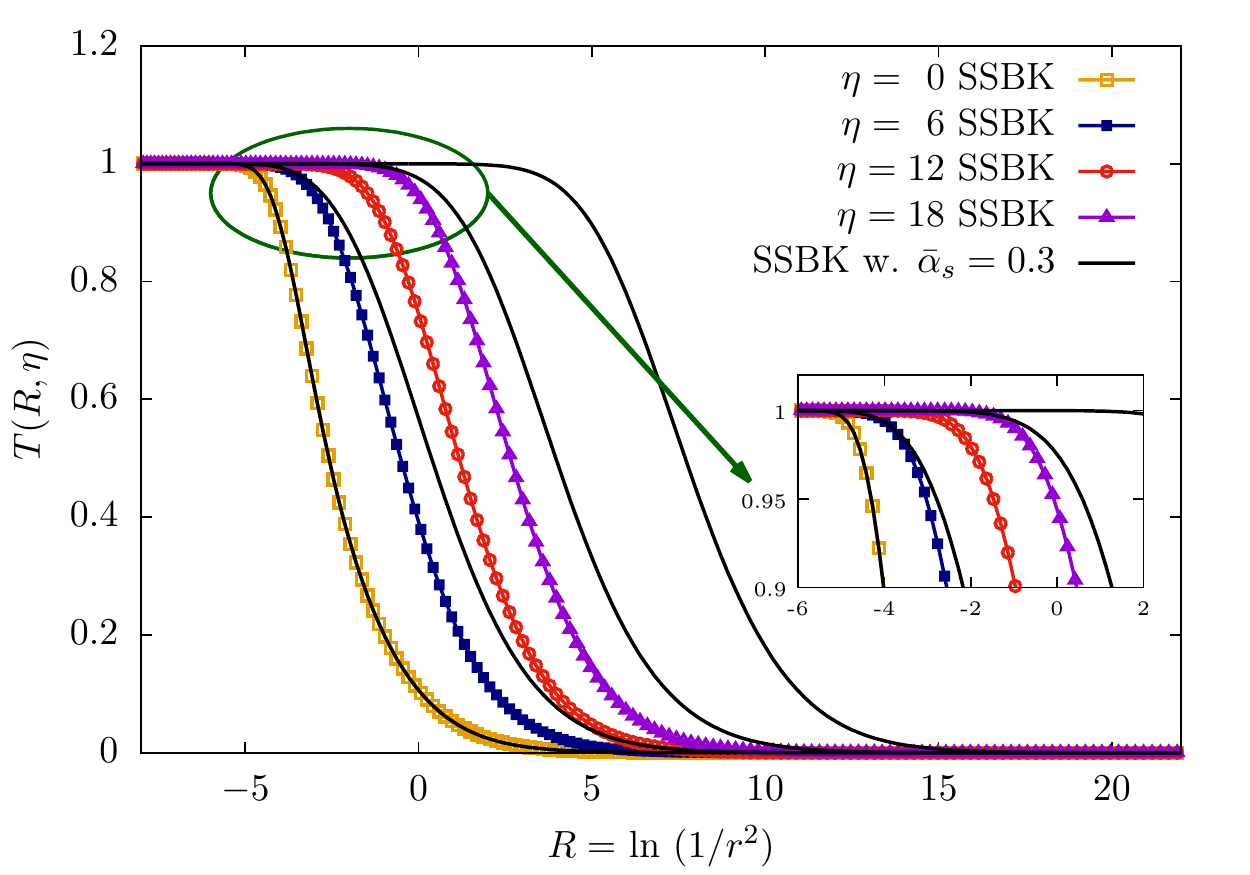, width=8cm,height=6cm}
\end{center}
\caption{The numerical solutions to non-local ciBK and SSBK evolution equations in $\eta$-representation for 4 different rapidities. The left hand panel gives the comparisons between the non-local ciBK and SSBK dipole amplitudes with the SDRCP. The right hand panel gives the comparisons between fixed coupling SSBK and running coupling SSBK dipole amplitudes. The inner diagrams are the zooming in amplitudes in the saturation region.}
\label{figSSBK}
\end{figure}

The left-hand panel of Fig.\ref{figSSBK} gives the solutions of the non-local ciBK and SSBK equations as a function of the dipole size in $\eta$-representation for 4 different rapidities. Note that we plot a zooming in diagram to clearly show the numerical results in the saturation region. One can see that the values of the SSBK dipole amplitudes are larger than the non-local ciBK ones for each corresponding rapidities in the saturation region. This outcome is consistent with the analytic findings, Eqs.(\ref{solSudBKrc}) and (\ref{solrcBK_eta}) in Sec.\ref{sec:BKinEta}, in which the linear rapidity dependence in the exponent of the $S$-matrix in non-local ciBK case is replaced by the rapidity raised to power of $3/2$ dependence due to the contribution of the sub-leading double logarithms. The right-hand panel of Fig.\ref{figSSBK} shows the solutions of the SSBK equation with the fixed coupling ($\abar=0.3$) and running coupling constant for 4 different rapidities. The zooming in diagram is used to assist in a clear view of the numerical results in the saturation region. We can see that the respective dipole amplitude with the the running coupling constant is smaller than the corresponding one with fixed coupling constant. This numerical result agrees with the analytic results, Eqs.(\ref{solSudBK}) and (\ref{solSudBKrc}), in which the quadratic rapidity dependence in the exponent of the $S$-matrix is replaced by the linear rapidity dependence once the SDRCP is used. The result also indicates that the running coupling effect takes a significant role in suppression of the evolution of the dipole amplitude in both $Y$-representation and $\eta$-representation.

\section{Confronting HERA data with Sudakov suppressed Balitsky-Kovchegov equation}
\label{sec:numsol}
In this section, we come to describe the HERA data\cite{H1ZEUS} for the inclusive DIS cross-section with the SSBK equation. The actual quantity we shall fit is the reduced $\gamma^\ast p$ cross-section which can be expressed in terms of the transverse, $\sigma_{\mathrm{T}}^{\gamma^\ast p}$, and longitudinal, $\sigma_{\mathrm{L}}^{\gamma^\ast p}$, cross-sections:
\be
\sigma_\mathrm{red} = \frac{Q^2}{4\pi^2\alpha_{\mathrm{em}}}\bigg[\sigma_{\mathrm{T}}^{\gamma^\ast p}+\frac{2(1-y)}{1+(1-y)^2}\sigma_{\mathrm{L}}^{\gamma^\ast p}\bigg],
\label{sigma-red}
\ee
with $y=Q^2/(sx)$ to be as the inelasticity variable and $s$ the squared center of mass collision energy. The transverse and longitudinal cross-sections in Eq.(\ref{sigma-red}) can be written as\cite{IMMST2,DIST20,Xiang07}
\be
\sigma_{\mathrm{T,L}}^{\gamma^\ast p} = \sum_f\int_0^1dz\int d^2r|\psi^{(f)}_\mathrm{T,L}(r,z;Q^2)|^2\sigma_\mathrm{dip}^{q\bar{q}}(r,x),
\label{sigma-TL}
\ee
where $|\psi^{(f)}_\mathrm{T,L}|^2$ is the squared light-cone wave function representing the probability for a virtual photon splitting into a quark-antiquark pair with flavor $f$, and can be written as
\be
|\psi^{(f)}_\mathrm{T}(r,z;Q^2)|^2 = e_q^2\frac{\alpha_\mathrm{em}N_c}{2\pi^2}\big\{\bar{Q}_f^2[z^2+(1-z)^2]K^2_1(r\bar{Q}_f)+m_f^2K_0^2(r\bar{Q}_f)\big\},
\label{wf-T}
\ee
\be
|\psi^{(f)}_\mathrm{L}(r,z;Q^2)|^2 = e_q^2\frac{\alpha_\mathrm{em}N_c}{2\pi^2}4Q^2z^2(1-z)^2K_0^2(r\bar{Q}_f),
\label{wf-L}
\ee
with $K_0$ and $K_1$ to be as the modified Bessel function of the second kind. The $\bar{Q}_f$ in Eqs.(\ref{wf-T}) and (\ref{wf-L}) is defined as $\bar{Q}_f^2=z(1-z)Q^2+m_f^2$ with $m_f$ to be as the quark mass. Note that we only use three light quarks with $m_{u,d,s}=140~\mathrm{MeV}$ in our fit.

The key ingredient in Eq.(\ref{sigma-TL}) is the dipole cross-section
\be
\sigma_\mathrm{dip} = \sigma_0 \big[1 - S(r, \eta)\big],
\label{sigma-dip}
\ee
which includes the most important information about the scattering between dipole and target. Here, the $S$-matrix is calculated by numerical solving the SSBK equation, and the $\sigma_0$ is viewed as a free parameter whose value is determined by fitting to the HERA data. We would like to point out that we use the one loop QCD coupling
\be
\abar(r^2) = \frac{1}{b\ln\big(\frac{4C^2}{r^2\Lambda^2}\big)}
\label{runningca}
\ee
\begin{table}[htbp]
  \begin{center}
  \begin{tabular}{cc|cccccc}
  \hline
  & QCD coupling &~~$\sigma_0$(mb)~~&~~$\bar{Q}_{s0}^2(\mathrm{GeV}^2)$~~&~~$\gamma$~~&~~$~~C^2$~~&~~$\chi^{2}/\mathrm{d.o.f}$ \\
  \hline
  & running $\alpha_s$    & 32.513   & 0.139   & 1.057  & 19.445  &  1.128     \\
  \hline
    \end{tabular}%
  \caption{Values of the fitting parameters and $\chi^{2}/\mathrm{d.o.f}$ from the fit to the reduced cross-section data points from \cite{H1ZEUS}.}
  \label{tabfit}
  \end{center}
\end{table}%
\begin{figure}[h!]
\setlength{\unitlength}{1.5cm}
\begin{center}
\epsfig{file=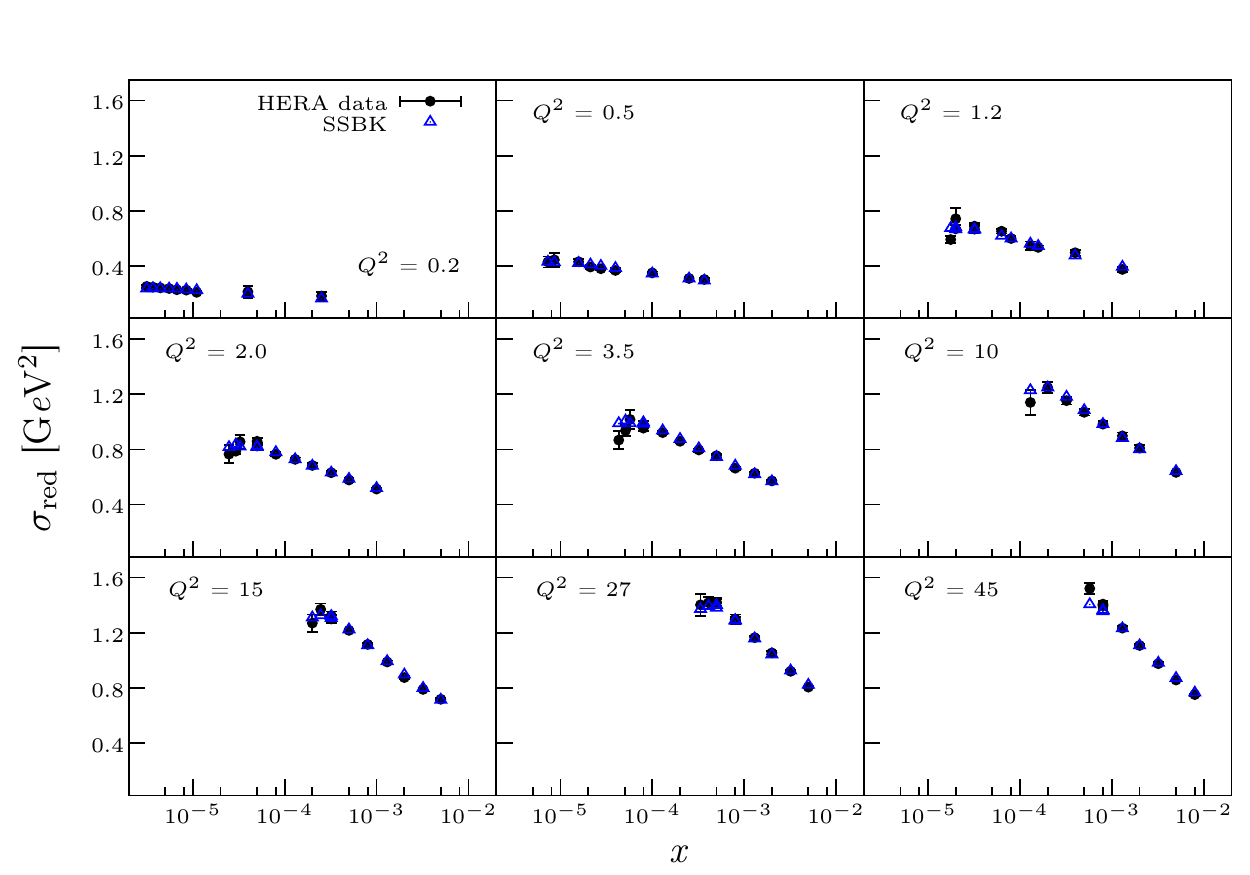, width=15cm,height=10cm}
\end{center}
\vspace{-6mm}
\caption{The reduced cross-section versus $x$ at different values of $Q^2$. The experimental data points are the combined measurement from H1 and ZEUS collaborations\cite{H1ZEUS}. We only plot the reduced cross-section for some typical values of $Q^2$. The others have the same good quality as well.}
\label{figred}
\end{figure}
\begin{figure}[h!]
\setlength{\unitlength}{1.5cm}
\begin{center}
\epsfig{file=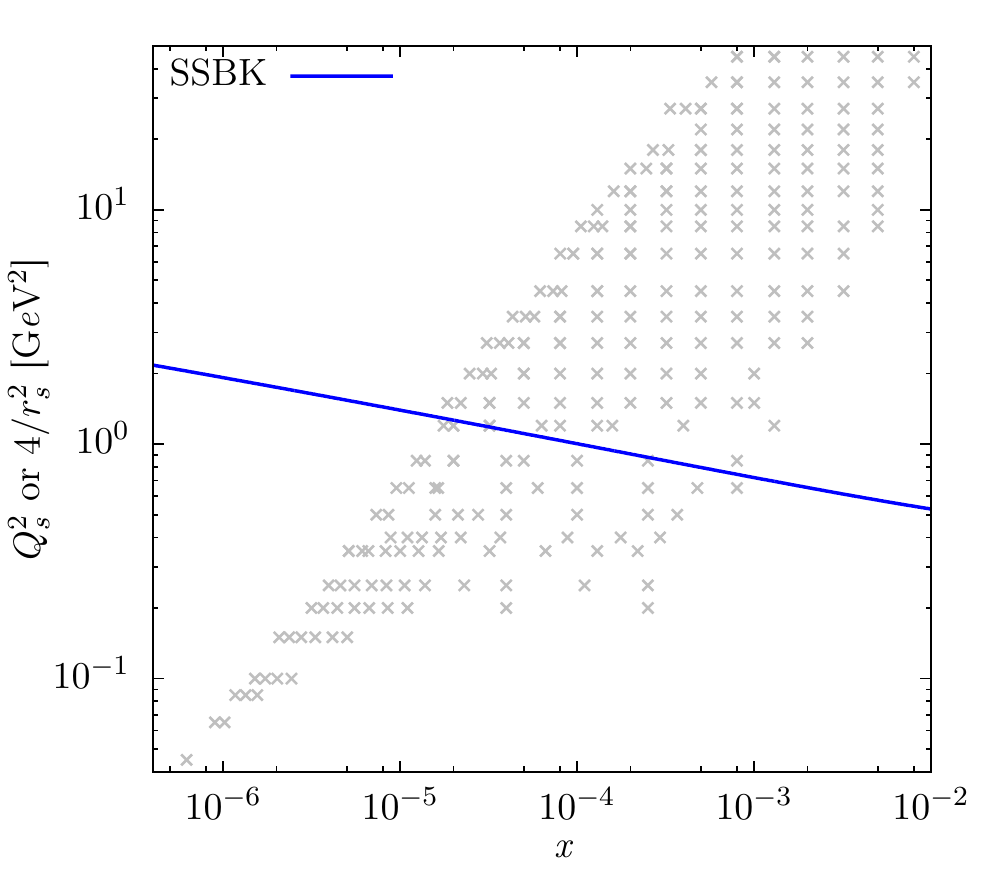, width=10cm,height=8cm}
\end{center}
\vspace{-6mm}
\caption{The squared saturation scale versus $x$, the $Q_s$ defined as $T(r=2/Q_s, \eta)=1/2$.}
\label{figqs}
\end{figure}
\noindent to fit the HERA data in term of the rather successful running coupling scheme in Refs.\cite{JS,JM}. We take $\Lambda=0.2~\mathrm{GeV}$ and treat the $C^2$ as a free parameter in our fit. In summary, we totally have four free parameters, $\sigma_0$, $\bar{Q}_{s0}^2$, $\gamma$ and $C^2$. The first one comes from the dipole cross-section, Eq.(\ref{sigma-dip}), the second and third parameters are from the initial condition, Eq.(\ref{IC}), and the final parameter originates from the QCD running coupling, Eq.(\ref{runningca}).

In the fit, we consider the combined data from HERA for the reduced cross-section in the kinematical range $x\leq 0.01$ and $0.045~\mathrm{GeV}^2<Q^2<50~\mathrm{GeV}^2$\cite{H1ZEUS}. The low limit on $Q^2$ is chosen low enough to justify the use of the BK dynamics rather than DGLAP evolution, and the upper limit large enough to include a large amount of perturbative data points. By using the criteria, we have 252 data points in our fit.

With the set up described above, we obtain rather good fits to the HERA data on the reduced cross-section. In Table \ref{tabfit}, we show the values of the free parameters. The reasonable value of $\chi^2/\mathrm{d.o.f}$ indicates that the SSBK equation with the SDRCP gives a rather successful description of the data. In Fig.\ref{figred}, we show the reduced cross-section as a function of the Bjorken variable $x$. Note that the Fig.\ref{figred} only plots the reduced cross-section for some typical values of $Q^2$, we have check that the others have the same good quality description as well. By comparison the data points (solid points in Fig.\ref{figred}) with the theoretical calculations (triangles in Fig.\ref{figred}), one can see that the values of the reduced cross-section calculated by the SSBK equation are in agreement with the HERA data points. This outcome is satisfying the theoretical expectation\cite{DIMST19}. To show the rapidity dependence of the saturation momentum, we extract the value of the $Q_s$ from the fit to the data via the definition $T(r=2/Q_s, \eta)=1/2$. In Fig.\ref{figqs}, we demonstrate the saturation momentum as a function of $x$. The $Q_s$ is shown on top of the data points that we use in the fit.


\begin{acknowledgments}
This work is supported by the National Natural Science Foundation of China under Grant Nos.11765005, 11305040, 11947119 and 11847152; the Fund of Science and Technology Department of Guizhou Province under Grant Nos.[2018]1023, and [2019]5653; the Education Department of Guizhou Province under Grant No.KY[2017]004; the National Key Research and Development Program of China under Grant No.2018YFE0104700, and Grant No.CCNU18ZDPY04.
\end{acknowledgments}



\begin{thebibliography}{99}

\bibitem{B} I.~Balitsky, {\it Nucl.\ Phys.}\ {\bf B463} (1996) 99.

\bibitem{JIMWLK1}
J.~Jalilian-Marian, A.~Kovner, A.~Leonidov, and H.~Weigert, {\it Nucl.\ Phys.}\ {\bf B504} (1997) 415.

\bibitem{JIMWLK2}
J.~Jalilian-Marian, A.~Kovner, A.~Leonidov, and H.~Weigert, {\it Phys. Rev.} {\bf D59} (1998) 014014.

\bibitem{JIMWLK3}
E.~Iancu, A.~Leonidov, and L.~McLerran, {\it Nucl.\ Phys.}\ {\bf A692} (2001) 583.

\bibitem{JIMWLK4}
E.~Ferreiro, E.~Iancu, A.~Leonidov, and L.~McLerran, {\it Nucl.\ Phys.}\ {\bf A703} (2002) 489.

\bibitem{K} Yu.V.~Kovchegov, {\it Phys. Rev.} {\bf D60} (1999) 034008; {\it ibid.} {\bf D61} (1999) 074018.

\bibitem{Xiang07} M.~Kozlov, A.~Shoshi, and W.~Xiang, {\it JHEP}\ {\bf 0710} (2007) 020.

\bibitem{JS} J.~Albacete, N.~Armesto, J.~Milhano, and C.~Salgado, {\it Phys. Rev.} {\bf D80} (2009) 034031.

\bibitem{JM} J.~Albacete, N.~Armesto, J.~Milhano, P.~Quiroga-Arias, and C.~Salgado, {\it Eur. Phys. J.} {\bf C71} (2011) 1705.

\bibitem{LM3} T.~Lappi, and H.~M$\mathrm{\ddot{a}}$ntysaari, {\it Phys. Rev.} {\bf D88} (2013) 114020.

\bibitem{CX} G.A.~Chirilli, B.W.~Xiao, and F.~Yuan, {\it Phys. Rev. Lett.} {\bf 108} (2012) 122301.

\bibitem{Albacete17} J.~Albacete, {\it Nucl.\ Phys.}\ {\bf A957} (2017) 71.

\bibitem{YCai20} Y.~Cai, W.~Xiang, M.~Wang, and D.~Zhou, {\it Chin.\ Phys.}\ {\bf C44} (2020) 074110.

\bibitem{Bnlo} I.~Balitsky, {\it Phys. Rev.} {\bf D75} (2007) 014001.

\bibitem{KW} Yu.V.~Kovchegov and H.~Weigert, {\it Nucl.\ Phys.}\ {\bf A784} (2007) 188.

\bibitem{BC08} I.~Balitsky, and G.~Chirilli, {\it Phys. Rev.} {\bf D77} (2008) 014019.

\bibitem{Beuf} G.~Beuf, {\it Phys. Rev.} {\bf D89} (2014) 074039.

\bibitem{NLOJIMWLK1} A.~Kovner, M.~Lublinsky, and Y.~Mulian, {\it Phys. Rev.} {\bf D89} (2014) 061704.

\bibitem{NLOJIMWLK2} A.~Kovner, M.~Lublinsky, and Y.~Mulian, {\it JHEP}\ {\bf 1404} (2014) 030.

\bibitem{NLOJIMWLK3} A.~Kovner, M.~Lublinsky, and Y.~Mulian, {\it JHEP}\ {\bf 1408} (2014) 114.

\bibitem{IMMST} E.~Iancu, J.~Madrigal, A.~Mueller, G.~Soyez, and D.~Triantafyllopoulos, {\it Phys.\ Lett.}\ {\bf B744} (2015) 293.

\bibitem{LM15} T.~Lappi, and H.~M$\mathrm{\ddot{a}}$ntysaari, {\it Phys. Rev.} {\bf D91} (2015) 074016.

\bibitem{LM16} T.~Lappi, and H.~M$\mathrm{\ddot{a}}$ntysaari, {\it Phys. Rev.} {\bf D93} (2016) 094004.

\bibitem{Levin16} C.~Contreras, E.~Levin, R.~Meneses, and I.~Potashnikova, {\it Phys. Rev.} {\bf D94} (2016) 114028.

\bibitem{Xiang17} W.~Xiang, S.~Cai, and D.~Zhou, {\it Phys. Rev.} {\bf D95} (2017) 116009.

\bibitem{DIMST19} B.~Ducloue, E.~Iancu, A.~Mueller, G.~Soyez, and D.~Triantafyllopoulos, {\it JHEP}\ {\bf 1904} (2019) 081.

\bibitem{JZhou19} D.~Zheng, and J.~Zhou, {\it JHEP}\ {\bf 1911} (2019) 177.

\bibitem{Xiang19} W.~Xiang, Y.~Cai, M.~Wang, and D.~Zhou, {\it Phys. Rev.} {\bf D99} (2019) 096026.

\bibitem{AK07} J.~Albacete, and Yu.V.~Kovchegov, {\it Phys. Rev.} {\bf D75} (2007) 125021.

\bibitem{Lipatov76} L.~Lipatov, {\it Sov. J. Nucl. Phys.} {\bf 23} (1976) 338.

\bibitem{KLF77} E.~Kuraev, L.~Lipatov, and V.~Fadin, {\it Sov. Phys. JETP} {\bf 45} (1977) 199.

\bibitem{BL78} I.~Balitsky, and L.~Lipatov, {\it Sov. J. Nucl. Phys.} {\bf 28} (1978) 822.

\bibitem{DIST20} B.~Ducloue, E.~Iancu, G.~Soyez, and D.~Triantafyllopoulos, {\it Phys.\ Lett.}\ {\bf B803} (2020) 135305.

\bibitem{Xiang20} W.~Xiang, Y.~Cai, M.~Wang, and D.~Zhou, {\it Phys. Rev.} {\bf D101} (2019) 076005.

\bibitem{Levin-Tuchin} E.~Levin, and K.~Tuchin, {\it Nucl.\ Phys.}\ {\bf B573} (2000) 83.

\bibitem{Mueller} A.~Mueller, hep-ph/0111244.

\bibitem{Albacete04} J.~Albacete, N.~Armesto, J.~Milhano, C.~Salgado and U.~Wiedemann, {\it Phys. Rev.} {\bf D71} (2004) 014003.

\bibitem{IMMST2}  E.~Iancu, J.~Madrigal, A.~Mueller, G.~Soyez, and D.~Triantafyllopoulos, {\it Phys.\ Lett.}\ {\bf B750} (2015) 643.

\bibitem{Cepila1} J.~Cepila, J.~Contreras, and M.~Matas, {\it Phys. Rev.} {\bf D99} (2019) 051052.

\bibitem{Xiang} W.~Xiang,  {\it Phys. Rev.} {\bf D79} (2009) 014012.

\bibitem{MV} L.~McLerran, and R.~Venugopalan, {\it Phys. Rev.} {\bf D49} (1994) 2233.

\bibitem{H1ZEUS} F.~Aaron, et al., {\it JHEP}\ {\bf 1001} (2010) 109.

\end{thebibliography}
\end{document}